\documentclass[aps,pra,twocolumn,showpacs,floatfix,10pt,superscriptaddress]{revtex4-1}
\pdfoutput=1

\usepackage{amsmath}
\usepackage{amssymb}
\usepackage{graphicx}

\begin{document}
\title{Photonic tuning of quasi-particle decay in a superfluid}
\author{G. K\'onya}
\affiliation{Institute for Solid State Physics and Optics, Wigner Research Centre, Hungarian Academy of Sciences, H-1525 Budapest P.O. Box 49, Hungary}
\author{G. Szirmai}
\affiliation{Institute for Solid State Physics and Optics, Wigner Research Centre, Hungarian Academy of Sciences, H-1525 Budapest P.O. Box 49, Hungary}
\author{D. Nagy}
\affiliation{Institute for Solid State Physics and Optics, Wigner Research Centre, Hungarian Academy of Sciences, H-1525 Budapest P.O. Box 49, Hungary}
\author{P. Domokos}
\affiliation{Institute for Solid State Physics and Optics, Wigner Research Centre, Hungarian Academy of Sciences, H-1525 Budapest P.O. Box 49, Hungary}

\begin{abstract}
We show that the damping rate of elementary excitations of hybrid systems close to a phase transition can undergo a remarkable resonance like enhancement before mode softening takes place. In particular, we consider the friction of a collective density wave in a homogeneous superfluid of weakly interacting bosonic atoms coupled to the electromagnetic field of a single mode optical resonator. Here the Beliaev damping can thus be controlled by an external laser drive and be enhanced by several orders of magnitude.    
\end{abstract}


\maketitle

Ultracold atoms coupled to the radiation field of an optical resonator form a long-range interacting many-body system \cite{ritsch2013cold,gopalakrishnan2009emergent,Strack2011Dicke,Jing2011Quantum} which proved to be suitable for the quantum simulation of the superradiant quantum phase transition of the Dicke-model \cite{nagy2010dicke,baumann2010dicke}.
Critical behaviour in non-equilibrium phase transitions between stationary phases of an open system  \cite{nagy2011critical,oztop2012excitations,dalla2013keldysh,Dimer2007Proposed,Morrison2008Dynamical,Diehl2010Dynamical,Horstmann2013Noise,Kessler2012Dissipative} cannot be cast in the usual formalism of the symmetry-breaking transition of the ground state. As it has been predicted \cite{nagy2011critical,oztop2012excitations} and recent experiments have approved \cite{Brennecke2013Realtime},  dissipation and the accompanying quantum fluctuations substantially modify the correlation functions and the critical exponents \cite{DallaTorre2010Quantum,dalla2013keldysh}. Dissipation is thus a key player in quantum criticality. One can ask the opposite: What is the effect of criticality on the dissipation channels? In general, the soft mode in phase transitions has to have a vanishing frequency and damping rate at the critical point, in accordance with the 'critical slowing down' phenomenon. In this Letter we show that, as the control parameter approaches the transition point, the damping rate of the soft-mode can undergo an unexpected, resonance-like, huge enhancement prior to vanishing.

Elementary excitations of a homogeneous Bose-Einstein condensate of ultracold atoms are collective density waves with different wave numbers which can be considered  ``quasi-particles". Besides the dispersion relation, the quasi-particles are characterized by a damping rate\cite{Jin1997Temperature,Chevy2002Transverse,Rowen2008Energy,Liu1997Theoretical,Giorgini1998Damping,Fedichev1998Damping}. The finite lifetime originates from two possible scattering processes among quasi-particles. The first one leads to Landau damping \cite{Guilleumas1999Temperature,Reidl2000Shifts,Jackson2003Landau}, where the selected excitation, together with another thermally excited one, merges into a third excitation of the system. This mechanism needs a thermal occupation of the other excitation, therefore it vanishes at zero temperature. In the second, so-called Beliaev damping process \cite{Hodby2001Experimental, Katz2002Beliaev}, the selected excitation decays directly into two lower energy excitations. This scattering process is the basic source of dissipation in a superfluid near zero temperature \cite{Kagan2001Damping}. In the following, we will consider the Landau and Beliaev collisional decay processes acting on quasi-particles when the homogeneous Bose gas is part of a bigger system which exhibits a phase transition.

\begin{figure}
\begin{center}
  \includegraphics{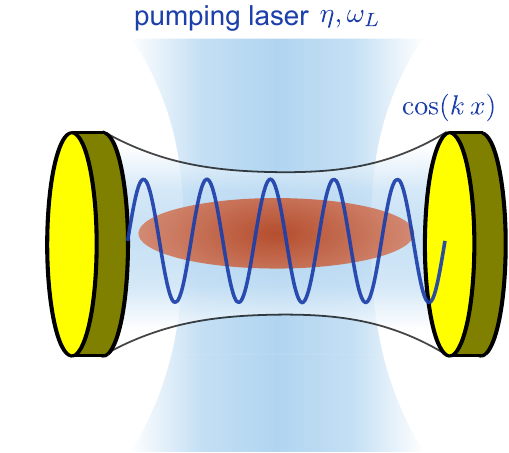}
  \caption{The scheme of the Fabry-P\'erot resonator sustaining an electromagnetic standing wave with a single $\cos(k\,x)$ mode function and containing a Bose-Einstein condensate of atoms illuminated from the side.}
  \label{fig:cavitybec}
\end{center}
\end{figure}

Suppose that a Bose-Einstein condensate of atoms is placed into an optical resonator \cite{brennecke2007cavity} and is illuminated by a coherent laser light from the side perpendicular to the cavity axis, see  Fig.~\ref{fig:cavitybec}. The laser frequency $\omega_L$ is far detuned from all atomic transitions, the absorption is thus negligibly small.  However, the driving laser is close to resonance with a single mode of the cavity, hence the atoms can efficiently scatter laser light into this mode. The photon scattering between the driving laser and the cavity is subject to interference in the many-particle system. The cavity mode function selects density-waves which are coupled by the collective scattering to the light field. Then the corresponding  quasi-particles are sensitive to  external control exerted by tuning the laser pump power or frequency. 

\begin{figure*}
\begin{center}
  \includegraphics{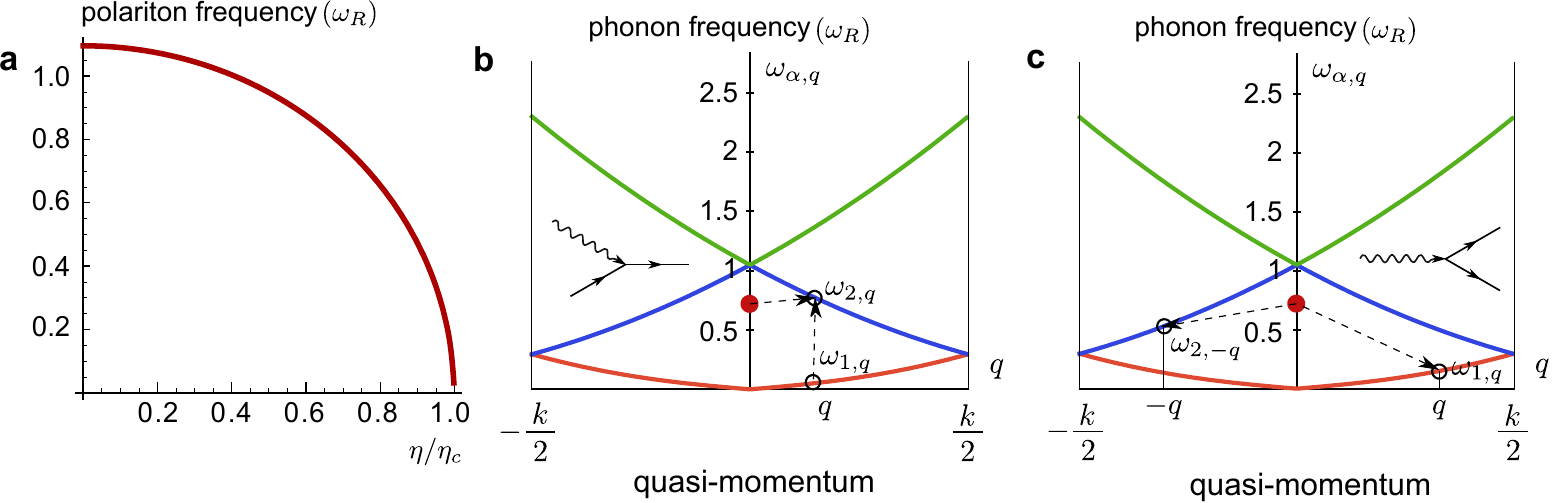}
  \caption{\textbf{Frequency schemes.} \textbf{a}, The frequency of the relevant density-wave quasi-particle dressed by photons, the polariton mode, as a function of the laser power. This is the soft mode of the normal-superradiant phase transition \cite{nagy2010dicke,mottl2012roton}, with vanishing frequency at the threshold of the superradiant phase. \textbf{b-c}, Phonon dispersion relations as a function of the quasi-momentum. \textbf{b}, Illustration of Landau damping. The polariton together with a phonon decays into a higher energy phonon. \textbf{c}, Illustration of Beliaev damping. The polariton decays into two lower energy phonons. In the damping processes energy and momentum are conserved.}
  \label{fig:frequencies}
\end{center}
\end{figure*}
At low temperature, the dilute gas of bosonic atoms with mass $m$, placed into a container with infinite length, is described by the Hamiltonian ($\hbar=1$): 
\begin{equation}
\label{eq:Hatom}
H_A=\int \hat{\Psi}^\dagger (x) \Biggl[ -\frac{1}{2m}\frac{d^2}{dx^2} + \frac{g}{2} \hat{\Psi}^\dagger(x) \hat{\Psi}(x)\Biggl]\hat{\Psi}(x)\,dx,
\end{equation}
where $g=4\pi a/(m w)$ is the strength of the low energy collisions with s-wave scattering length $a$. For simplicity we consider only the one-dimensional motion of the atoms along the cavity axis, inside an elongated trap with the transverse size of the condensate taken to be $w$. The entire Hamiltonian is then $H=H_A+H_C$, with
\begin{multline}
\label{eq:Hcavity}
H_C= -\Delta_C \hat{a}^\dagger\hat{a}+ \int_0^L \hat{\Psi}^\dag (x) \Big[\eta \left( \hat{a}^\dag +\hat{a} \right) \cos(k x)\\
+ U_0 \, \hat{a}^\dagger\hat{a} \, \cos^2 (k x) \Big] \hat{\Psi}(x) \, dx.
\end{multline}
The first term describes the bare cavity energy in a frame rotating with the pumping laser frequency, $\Delta_C=\omega_L-\omega_C$. Terms in the integral represent two kinds of optical processes involving the atom cloud density. The first one is a laser drive of the cavity via photon scattering off the atoms. The corresponding effective amplitude is $\eta=\Omega g_0/\Delta_A$, with $\Omega$ the Rabi frequency of the pumping laser, $g_0$ the single-photon Rabi frequency in the cavity, $\Delta_A=\omega_L-\omega_A$ the detuning of the laser from the atomic resonance. The cavity mode function is $\cos(kx)$, with wavenumber $k$. The second interaction term  having $a^\dagger a$ is the absorption of cavity photons and induced emission  back into the cavity \cite{Murch2008Observation,Wolke2012Cavity}. This coherent scattering forms an optical lattice potential $\cos^2(k x)$  for the atomic matter wave with a depth proportional to the intensity with a coefficient $U_0=g_0^2/\Delta_A$. 

We briefly recall here that the system defined by this Hamiltonian admits a very simple solution for weak driving strength, which is referred to as the normal phase.  When the density of the atom cloud is constant along the cavity axis, the effective driving term vanishes by integrating out the $\cos(k x)$ mode function over the condensate.  This means destructive interference in the scattering, and no photon field builds up in the cavity. Nothing modulates then the quasi-homogeneous condensate density. This solution breaks down above a certain pump power $\eta_c=\sqrt{-\Delta_C\, \omega_R}$  \cite{nagy2010dicke,baumann2010dicke,baumann2011exploring}, where $\omega_R=\hbar k^2/2m$ is the recoil frequency. Above the critical point $\eta>\eta_c$ a stable periodic modulation of the atomic density is formed. However, in the following we consider only below-threshold driving strength with the corresponding homogeneous superfluid state.

The periodicity of the interaction terms with the cavity wavelength suggests the decomposition of the matter wave field in terms of Bloch states as
\begin{equation}
\hat{\Psi}(x) = \frac{1}{\sqrt{L}} \sum_q e^{i q x} \Big[ \hat{b}_q + \sqrt{2} \, \cos(k x) \, \hat{c}_q + \sqrt{2} \sin(k x) \hat{s}_q \Big],
\end{equation}
where $\hat b_q$, $\hat c_q$, and $\hat s_q$ are annihilation operators of atomic single particle states with wavefunctions $e^{iqx}$, $\cos(k x)e^{iqx}$, and $\sin(kx)e^{iqx}$, respectively. Here we introduced a quasi-momentum $q\in[-k/2,k/2]$. We consider only the three lowest energy bands, higher order harmonics of the type $\cos(n k x)$ ($n=2,3,..$) are not relevant for the present study \cite{konya2011}. 

All the boson mode operators can be split to mean field and fluctuation parts. In the present geometry, for weak driving only the homogeneous atom field mode, which contains the condensate, has non-vanishing mean $\hat{b}_0 \rightarrow\sqrt{N_c} + \hat{b}_0$, where $N_c$ is the number of condensate atoms. All the other excitation modes as well as the photon field have zero mean amplitude.  The Heisenberg equations of motion for the fluctuations can be cast into the form
of a hierarchy of terms with different powers of the condensate atom number $\sqrt{N_c}$, 
\begin{subequations}
 \label{eq:CoupledLangevin}
\begin{multline}
\label{eq:poleq}
i \frac{d}{dt} \, {v}_\mu = \sum_{\nu} F_{\mu \nu} \, {v}_{\nu}  + \frac{1}{\sqrt{N_c}} \sum_q \sum_{\alpha , \beta} V^{\alpha \beta}_{\mu} \\
\times\left[ {{w}^\dag_{\alpha} (q) {w}_{\beta} (q)} - \left\langle {w}^\dag_{\alpha} (q) {w}_{\beta} (q) \right\rangle \right] \,,
\end{multline}
\begin{equation}
i \frac{d}{dt} \, {w}_\mu (q) = \sum_{\nu} G_{\mu \nu} (q) \, {w}_{\nu} (q)
+ \frac{1}{\sqrt{N_c}} \sum_{\alpha , \beta} \, W^{\alpha \beta}_{\mu} \; { {v}_{\alpha} \, {w}_{\beta}(q)}\,,
\end{equation}
\end{subequations}
where we used the compact vector notation 
\begin{subequations}
\begin{align}
{v}  &=\left( {a} \, , \, {a}^\dag \, , \,  {b}_0 \, , \, {b}_0^\dag \, , \, {{c}_0} \, , \, {{c}_0^\dag} , \, {{s}_0} \, , \, {{s}_0^\dag}  \right)^T, \label{eq:polariton}\\
{w}(q) &=\left( {b}_q \, , \, {b}_{-q}^\dag \, , \, {c}_q \, , \, {c}_{-q}^\dag \, , \, {s}_q \, , \, {s}_{-q}^\dag \right)^T, \label{eq:phonon}
\end{align}
\end{subequations}
for the $q=0$ and $q\neq0$ modes, respectively. Orders with higher powers of  $1/\sqrt{N_c}$ are omitted. 

The highest order describes a linear coupling between the modes, which corresponds to the Bogoliubov approach. Eigenmodes of the linear system define the quasi-particles. Up to this order, modes with different quasi-momentum magnitude $|q|$ do not couple.  Moreover, only the $q=0$ modes couple to the photon degree of freedom by the laser-induced interaction  \eqref{eq:Hcavity}. Therefore,  the set of modes $\left\{a, b_0, c_0, s_0\right\}$, gathered in $v$ in Eq.~(\ref{eq:polariton}),  form ``polariton" modes and has to be treated separately.  Of special importance is the excitation mode $c_0$ which matches exactly the cavity mode function $\cos(k x)$ and hence can be populated directly from the homogeneous BEC mode $b_0$ by scattering photons between the laser and the cavity mode.  The $q\neq0$ quasi-momentum excitations form the familiar Bogoliubov spectrum of the homogeneous BEC, represented by the dispersion curves in Fig.~\ref{fig:frequencies}b,c, and are referred  to as ``phonons'' in the following. 

Beyond the standard Bogoliubov approximation, the next order accounts for the interactions between quasi-particles, in particular, the cross-coupling between polaritons and phonons. In Eq.~(\ref{eq:CoupledLangevin}), one polariton mode is coupled to two phonons, which is in accordance with the scattering processes underlying the Landau and Beliaev damping, as sketched in Fig.~\ref{fig:frequencies}. Assuming large condensate size, the phonons are spectrally dense and form a dissipation bath for the quasi-particles, including the phonons themselves.  The damping is thus an intrinsic property originating from the short-range s-wave scattering. The rate of damping of a given polariton mode can be calculated within the Markov approximation \cite{Graham1999Langevin} which relies on that the phonons span a broad frequency range compared to the decay rate \footnote{When integrating out the phonon variables, we adopted the three-dimensional density of phonon modes to better mimic the real experimental situation.}.

\begin{figure}[htb]
\begin{center}
  \includegraphics{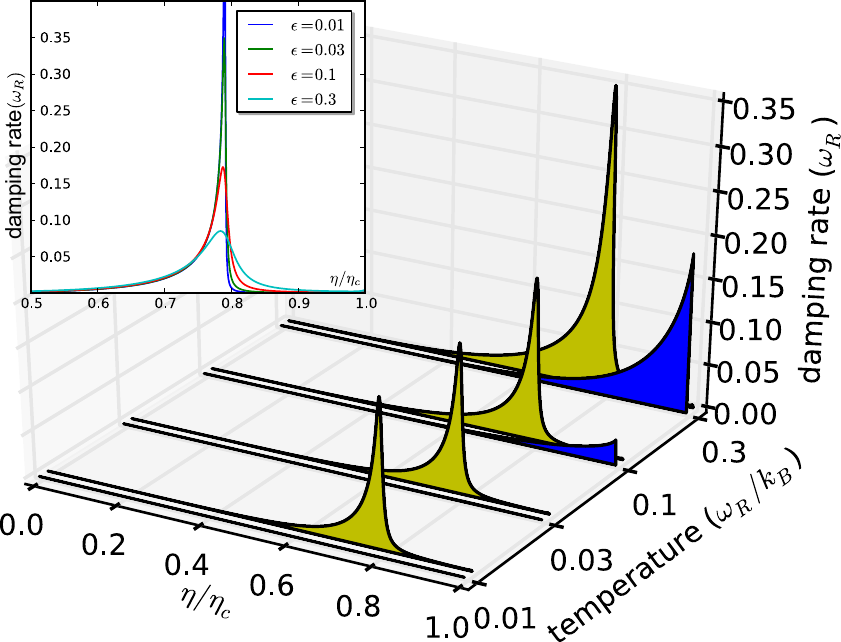}
  \caption{Landau (blue) and Beliaev (yellow) damping rates are plotted as a function of the pumping strength $\eta$ at various temperatures. The Landau damping rate increases towards the critical point but is suppressed (invisible) for temperatures below $k_BT \sim 0.1 \hbar \omega_R$. The inset shows  the strong peak of the Beliaev damping (at $k_B\,T=0.01$, it hardly depends on temperature)  for various values of the phenomenological parameter $\epsilon$ accounting for the decay rates of the phonons. The parameters are  $N_c=10^4$, $k\, L /(2\pi)=1000$, $N_c g/L = 0.1 \omega_R$, $\Delta_C=-1000 \omega_R$.}
  \label{fig:gamma}
\end{center}
\end{figure}

Figure~\ref{fig:gamma} shows the Landau and the Beliaev damping rates, separately, of the polariton which is composed dominantly of the $c_0$ mode. In the considered geometry, this quasi-particle is the most susceptible to the external control parameter $\eta$ which can be varied either by the pump laser power or by its detuning $\Delta_A$ from the atomic resonance.
The damping rate  starts from the value characteristic of this excitation in free space, then the Beliaev part develops a strong resonance peak at $\eta/\eta_c  \approx 0.8$. Further increasing the control parameter, the damping rate falls down abruptly. 

The tunability of the damping of a quasi-particle is due to the dressing by cavity photons, and the broad range of tunability is due to the criticality in the system. The frequency of one of the polariton modes [resulting from the diagonalization of $F_{\mu\nu}$ in Eq.~\eqref{eq:poleq}]  depends significantly on the interaction strength, as presented in Fig. \ref{fig:frequencies}a, since this is the soft mode of the normal-superradiant transition. For $\eta=0$ the polariton mode frequency is just at the point where the second and third phonon bands touch  (see Figs.~\ref{fig:frequencies}b,c). When the laser is turned on, for increasing $\eta$ this particular point of the excitation branch departs from the dispersion curve and its frequency gradually decreases according to Fig. \ref{fig:frequencies}a. For energy conservation, varying the polariton frequency amounts to sampling the bath at different points of the spectral density function. The resulting damping rates are thus tuned by the control parameter $\eta$. 

The key to understand the resonant behaviour in the superfluid at zero temperature is that the spectral density function in this case is not directly the Bogoliubov phonon spectrum. In a Beliaev type decay process, the energy of the polariton is distributed between the two phonons interlinked by momentum and energy conservation. From the former it follows that the phonons have to have opposite quasi-momenta and one has to be from the first and the other one from the second band, as illustrated in Fig.~\ref{fig:frequencies}c. An effective spectral density function can be derived for such a third-order decay process. In particular, for polariton frequencies at about $\omega_R/2$ the polariton decays into two phonons being at the opposite edges of the Brillouin zone $|q|\approx k/2$. Here the dispersion relation is linear hence a continuum set of pairs $+q \lesssim k/2$ on the lower branch and $-k/2 \lesssim -q$ on the upper branch fulfills both the momentum and energy conservation laws. This yields a diverging effective spectral density, and ultimately, this is the underlying reason of the peak in $\gamma$ at $\eta/\eta_c=0.8$.  Below this polariton frequency, the decay process becomes necessarily non-resonant and more and more suppressed.  All this analysis is valid up to the point $\eta/\eta_c =1$, since this is a critical point where the homogeneous mean field solution collapses. 

We introduced the phenomenological parameter $\epsilon$ which accounts for the summed damping rates of the phonons taking part in the decay process. The usual Landau and Beliaev formulae should be retained for the  $q\neq0$ phonon modes, as they not couple to the photon field.  In the present model, we simply used a single (fitting) parameter $\epsilon$ instead of treating $\epsilon$ as a function of the quasi-momenta and calculating it from the microscopic model. The dependence of the quasi-particle damping rate on the $\epsilon$ phonon damping rate is displayed in the inset of  Fig.~\ref{fig:gamma}.  The phonon decay is typically in the range of a few hundreds Hz, therefore we expect the curve associated with $\epsilon=0.1$ to be the best prediction to experimental data.

Recent experiments performed on this system found such a peak in the decay rate of the $\cos{k x}$ excitation mode, see Fig.~4 in \cite{Brennecke2013Realtime}. Our theory reveals then that the observed peak can be a manifestation of the Beliaev damping which is typically a negligible process but here the light shift of the polariton frequency leads to a significant enhancement. Other effects, such as the finite size effect can also contribute \cite{Kulkarni2013Cavity} or modify the peak. Especially, the peak can be broadened by an external trapping which might open a non-negligible gap in the dispersion relation at the edges of the Brillouin zone.  We emphasize,  however, that the photon-assisted Beliaev damping effect described above is an intrinsic property of the infinite ultracold atom gas system. It exists in the thermodynamic limit defined by the length $L\rightarrow \infty$ and the number of atoms $N\rightarrow\infty$ such that the density $N/L$ is constant.  Based on our detailed calculation \cite{KonyaLong} we expect that the damping rate is proportional to the density $N_c/L$.

We thank Ferdinand Brennecke, Rafael Mottl and Peter Sz\'epfalusy for discussions. This work was supported by the Hungarian National Office for Research and Technology under the contract ERC\_HU\_09 OPTOMECH, the Hungarian Academy of Sciences (Lend\"ulet Program, LP2011-016), and the Hungarian Scientific Research Fund (grant no. PD104652). G.Sz. also acknowledges support from the J\'anos Bolyai Scholarship.

\bibliography{photongrease}

\end{document}